\documentclass[pra
	       ,nofootinbib
	       ,floatfix
	       ,superscriptaddress
	       ,twocolumn
	       ]{revtex4-2}
	       
\usepackage{amsmath} % include math package
\usepackage{graphicx} % include graphics package	
\usepackage{color} % for colored text
\usepackage[utf8]{inputenc}
\usepackage{siunitx}
\usepackage[hidelinks]{hyperref}

\begin{document} 

\title{Non-invasive chemically selective energy delivery and focusing inside a scattering medium guided by Raman scattering}

\author{Bingxin~Tian}
\affiliation{School of Optoelectronic Engineering, Xi'an Technological University, No.2 Xuefuzhonglu Road, Weiyang District, Xi'an City, Shaanxi Province, 710021, People's Republic of China}

\author{Bernhard~Rauer}
\affiliation{Laboratoire Kastler Brossel, ENS-Université PSL, CNRS, Sorbonne Université, Collège de France, 24 rue Lhomond, 75005 Paris, France}

\author{Antoine~Boniface}
\affiliation{Laboratoire Kastler Brossel, ENS-Université PSL, CNRS, Sorbonne Université, Collège de France, 24 rue Lhomond, 75005 Paris, France}

\author{Jun~Han}
\affiliation{School of Optoelectronic Engineering, Xi'an Technological University, No.2 Xuefuzhonglu Road, Weiyang District, Xi'an City, Shaanxi Province, 710021, People's Republic of China}

\author{Sylvain~Gigan}
\affiliation{Laboratoire Kastler Brossel, ENS-Université PSL, CNRS, Sorbonne Université, Collège de France, 24 rue Lhomond, 75005 Paris, France}

\author{Hilton B. de Aguiar}
\email{h.aguiar@lkb.ens.fr}
\affiliation{Laboratoire Kastler Brossel, ENS-Université PSL, CNRS, Sorbonne Université, Collège de France, 24 rue Lhomond, 75005 Paris, France}

\date{\today}

\begin{abstract}
Raman scattering is a chemically selective probing mechanism with diverse applications in industry and clinical settings. Yet, most samples are optically opaque limiting the applicability of Raman probing at depth. Here, we demonstrate chemically selective energy deposition behind a scattering medium by combining prior information on the chemical's spectrum with the measurement of a spectrally resolved Raman speckle as a feedback mechanism for wavefront shaping. We demonstrate unprecedented six-fold signal enhancement in an epi-geometry, realizing targeted energy deposition and focusing on selected Raman active particles.
\end{abstract}

\maketitle

In order to avoid exogenous labelling and increase chemical selectivity, Raman based technologies have emerged as powerful tools for biomedical applications. For instance, in the last decades, intensive efforts have been put in developing novel surface-enhanced Raman spectroscopy (SERS) reporters~\cite{Tabish2020}, deep-sensing techniques such as spatially-offset Raman spectroscopy (SORS)~\cite{mosca2021spatially}, and fast imaging coherent Raman scattering microscopy techniques~\cite{Cheng2015}. However, these technologies are limited by a compromise in probing depth and spatial resolution. The main nuisance in any visible-light-based approach is light scattering and optical aberrations of biological specimens \cite{1997Measurement,2004Characterizing}.

Several \textit{spectroscopy} applications would benefit from a chemically-selective energy deposition deep inside scattering media. For instance, bright Raman-guided signals (SERS process) are used in photodynamic therapy to locally deposit energy through efficient light absorbers~\cite{Tabish2020} or for detection of pathologies using targets with chemical specificity through their Raman spectra~\cite{Noonan2018,Mcqueenie2012,Harmsen2015}, to cite a few applications where high-resolution imaging is not the main objective. Since Raman guidestars can be spatially heterogeneous, finding a mechanism that uses Raman readout to locally deliver more efficient energy would be highly beneficial as it reduces the excitation power to levels below authorized clinical standards. Increasing the efficiency of energy deposition would also allow to increase the probing penetration depth of the method.

Various recent attempts have been made to increase the penetration depth of Raman \textit{microscopy}. 
A particularly challenging task is to push the probing depth beyond the limits of ballistic light transmission using methods from adaptive optics~\cite{Wright2007}. Wavefront shaping techniques that manipulate light beyond lower order aberrations have shown focusing and image transmission through complex media~\cite{Vellekoop2007,Popoff2010}, hence bringing a potential solution to increase the penetration depth in high resolution microscopy~\cite{Gigan2021}. Toward translation of these techniques to Raman imaging, some progress has been made~\cite{Gusachenko2017,Tragardh2019,Amitonova2020,Hofer2020,Paniagua-Diaz2018} in geometries that allow physical access to the target image plane, a setting typically found in fiber-based endoscopy.
However, translating these outcomes to a non-invasive Raman imaging setting has been elusive as most biomedical Raman probing applications rely on epi-detection.

Non-invasive focusing within scattering environments requires guidestars: reporters from within the sample that provide a feedback signal for wavefront shaping optimization~\cite{Horstmeyer2015}. Typical candidates for feedback mechanisms are non-linear optical processes~\cite{Katz2014b}, as they guarantee focusing in a single target for non-sparse samples. Notable exceptions exist using wavefront shaping with a linear feedback mechanism~\cite{Thompson2016,Vellekoop2008}, however using highly sparse samples which are not typical of Raman-active media. The fundamental issue with linear optical feedback mechanisms (i.e. linear fluorescence and spontaneous Raman processes) is that there is no difference in the total signal upon focusing on a single target or distributing the power over all available targets. Therefore, linear feedback optimization routines can only converges to a single focus in samples that are highly sparse. 
However, recent approaches imaging the spatial speckled distribution of the emitted signal, rather than just the integrated total signal, were able to circumvent this limitation.
Optimising the signal speckle's variance, a non-linear metric, enabled non-invasive focusing for linear fluorescence in non-sparse samples~\cite{Boniface2019,Anat2019Light,Li2020}. 

Yet, extending wavefront shaping methods to Raman is far from trivial. Different from fluorescence where labelling can be controlled and only a targeted plane is addressed, in Raman the surrounding medium also generates a strong signal that buries the target response. In particular, the shot-noise of this background will complicate the evaluation of the target speckle.

Here, we present an approach that uses a priori chemical information and feedback-based wavefront shaping allowing for highly selective energy deposition. We use an imaging spectrometer to probe both the spatial and spectral distribution of the emitted Raman signal, in conjunction with the known spectra, to continuously optimise the incident wavefront. 
By that, we realise chemically selective energy delivery with high chemical specificity, focusing on single particle given the sample is sparse enough.

\begin{figure}[t]
\centering\includegraphics[width=1.0\columnwidth]{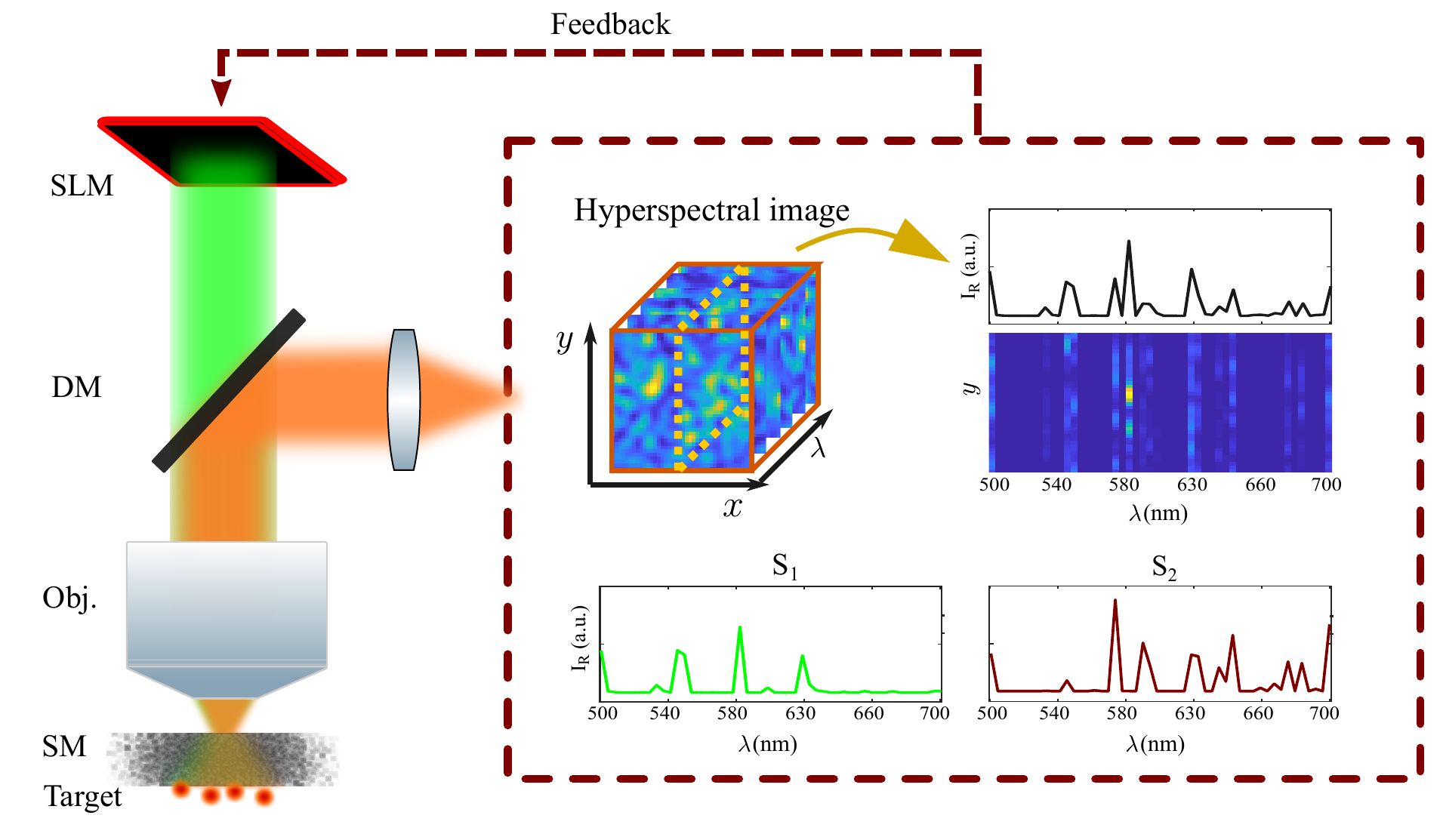}
\caption{Principle for non-invasive chemically-selective energy deposition guided by a Raman signal. A simplified optical layout of the experiment is shown on the left. The targets, placed behind the scattering medium (SM), are illuminated by a laser source modulated with a spatial light modulator (SLM). The excited Raman light is scattered back, collected by an objective (Obj) and reflected off a dichroic mirror (DM) towards the entrance of an imaging spectrometer that spectrally resolves the speckle pattern. 
A visualization of the recorded data is shown in the dashed box on the right.
The hyperspectral image is determined by the scattering and the target spectra, $S_1$ and $S_2$.
The transmission of the scattering media is spectrally fully decorrelated leading to speckle varying in $x$, $y$ and $\lambda$.
Using the a priori chemical information about the targets, one can design feedback mechanisms allowing for single-target focusing or targeted energy deposition inside the scattering medium.}
\label{fig:principle}
\end{figure}

We start by considering a simulation of a light delivery in a chemically complex sample where the Raman spectra of the targets are well known. 
The wavefront of the excitation light is actively changed mode by mode with a spatial light modulator (SLM) in order to maximize a suitable metric (see Fig. \ref{fig:principle}). 
We model a complex medium by a fully random transmission matrix without any memory effect and place $N_t$ target particles behind it. 
Upon excitation of multiple targets, the broadband Raman response of each source is transmitted back through the same scattering medium, however with spectrally decorrelated transmission matrices. 
The targets' response adds up incoherently at the camera of the imaging spectrometer outside the scattering medium (Fig. \ref{fig:principle}, hyperspectral image). 
Here, we consider two spectrally distinct chemicals contributing with comparable weights. 
While we model two relatively independent spectra (with random spectral peak positions), in reality there is always a background environment that generates a signal from $N_b\gg N_t$ sources. 
For this reason, any optimization algorithm has an initialization problem as the background Raman response buries the initial speckle signal of the targets.

\begin{figure}[t]
\centering\includegraphics[width=1.0\columnwidth]{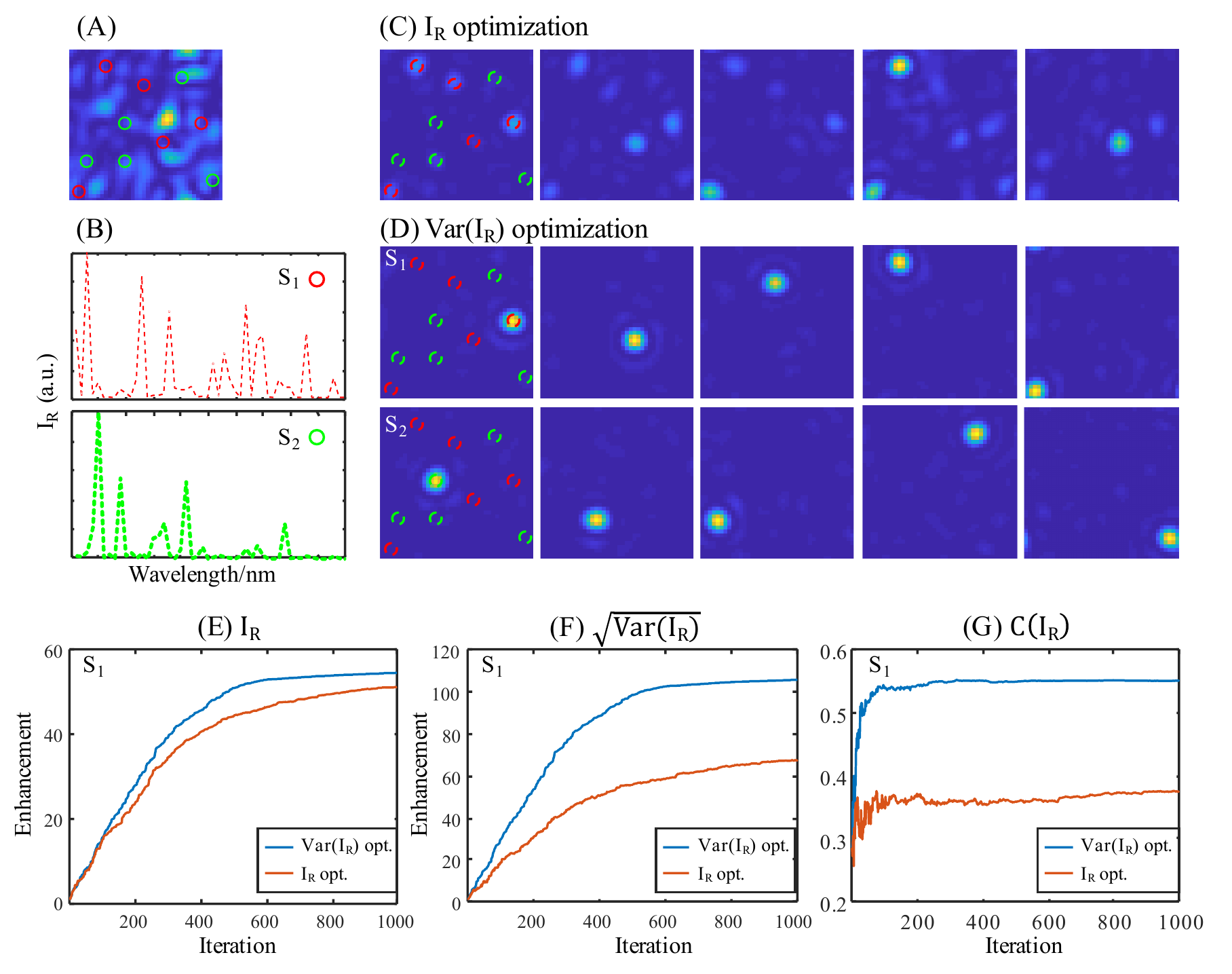}
\caption{Simulation results. (A) Targets of two chemicals behind the scattering medium are illuminated by a random pattern. 
Their location is indicated by the red and green circles, respectively.
(B) Corresponding spectra, modeled by random peak positions.
(C) Excitation patterns resulting from an intensity optimization for five different scattering realization but the same target distribution.
(D) Same for the variance optimization, targeting the two distinct spectra.
On which particle of one chemical the focus converges depends on the speckle realization.
(E-F) Enhancement of the total intensity and the variance of the Raman signal, respectively.
Comparing the two optimization metrics shows little difference in case of the intensity enhancement while the variance enhancement is substantially larger in the case of the variance optimization.
(G) Evolution of the speckle contrast for both intensity and variance optimization. 
Here, the spectrally weighted speckle was first summed before the contrast was evaluated, leading to the final values below 1 for the variance optimization.
For each optimization, there were 256 modes controlled on the SLM and each was optimised 4 times.
}
\label{fig:simulation}
\end{figure}

In this simulation, we now compare two possible metrics for the wavefront optimization: the total intensity and the spatial variance of the epi-detected Raman signal. 
In both cases, we record the spectrally-resolved speckle of multiple buried targets and weight it according to their known spectra (Fig.~\ref{fig:simulation}A and B).
For the intensity-based feedback we simply sum the weighted response over $x$, $y$ and $\lambda$ and optimise the input wavefront to maximise this quantity.
In this case, the excitation light converges on a single chemical but not on a single target (Fig.~\ref{fig:simulation}C).
This is due to the linear dependence of the Raman signal on the excitation light, where illuminating a single or multiple targets with a fixed total power will generate the same output signal.
In contrast, tuning the input wavefront to optimise the spatial variance of the weighted speckle pattern and summing it over $\lambda$ leads to a reliable focus on a single target of the desired chemical (Fig.~\ref{fig:simulation}D).
The reason for this is that the variance depends both on the intensity and contrast of the detected Raman speckle 
\begin{equation}
\mathrm{Var}(I_R) = \big( C(I_R) \cdot \langle I_R \rangle \big)^2,
\end{equation}
with the contrast being defined as $C = \sqrt{\langle I_R^2 \rangle / \langle I_R \rangle^2 - 1}$.
As the Raman speckle is the incoherent sum of the signals emitted by all targets, maximising its contrast facilitates convergence on a single target.
The additional intensity dependence further assures that also the total energy deposited on that target is maximised.
Therefore, while the enhancement of the intensity is similar in for both optimization metrics (Fig.~\ref{fig:simulation}E), the variance and the contrast improve more in the case of the variance feedback (Fig.~\ref{fig:simulation}F and G). 
Finally, an aspect that is particularly important for spontaneous Raman scattering is the typically low photon-budget.
We thus studied the effect of noise and found, despite a decrease in the single-target focusing performance, the chemical selectivity to be robust. 
This is particularly interesting for settings where the main interest lies in energy deposition, rather than imaging.

To experimentally demonstrate this chemically selective technique, we devise a sample containing two components.
Few Raman active particles (diamonds $\sim$\SI{1}{\micro\meter} in size) are placed on a glass cover slip.
The glass gives a background signal while the particles act as the targets.
The bottom surface of the cover slip facing the objective was sand blasted to introduce scattering (see Supplemental Document S2). 
We use a \SI{532}{\nano\meter} continuous-wave laser source whose wavefront is modulated by a phase-only SLM before illuminating the sample.
The emitted Raman light propagates backwards through the scattering layer and the illumination objective before it is epi-detected on an imaging spectrometer. 
A slit in the spectrometer selects only a slice of the scattered Raman image such that we measure a spectrally resolved one-dimensional speckle.
An additional objective in the transmission path images the sample and the excitation speckle, but is used only for monitoring (for details see Supplemental Document S1).

For the wavefront optimization we choose a continuous sequential algorithm, modulating all of the excitation light.
In each iteration half of SLM pixels are phase-stepped between 0 and 2$\pi$ while the other half is kept constant and acts as a reference.
The active pixels are then set to the value at which the chosen metric is maximised. 
The selection of active pixels follows a sequence of Hadamard patterns, defining a complete basis of the modulated wavefront.
For the experiments presented here we loop through this basis two times.
As a metric we use either the spatial variance or the total intensity of the speckle recorded at the Raman shift \SI{1332}{\per\centi\meter} associated with diamond.
For this we first subtract the background estimated from the signal recorded at wavelengths in the vicinity of the diamond resonance.

\begin{figure}[t]
\centering\includegraphics[width=1.0\columnwidth]{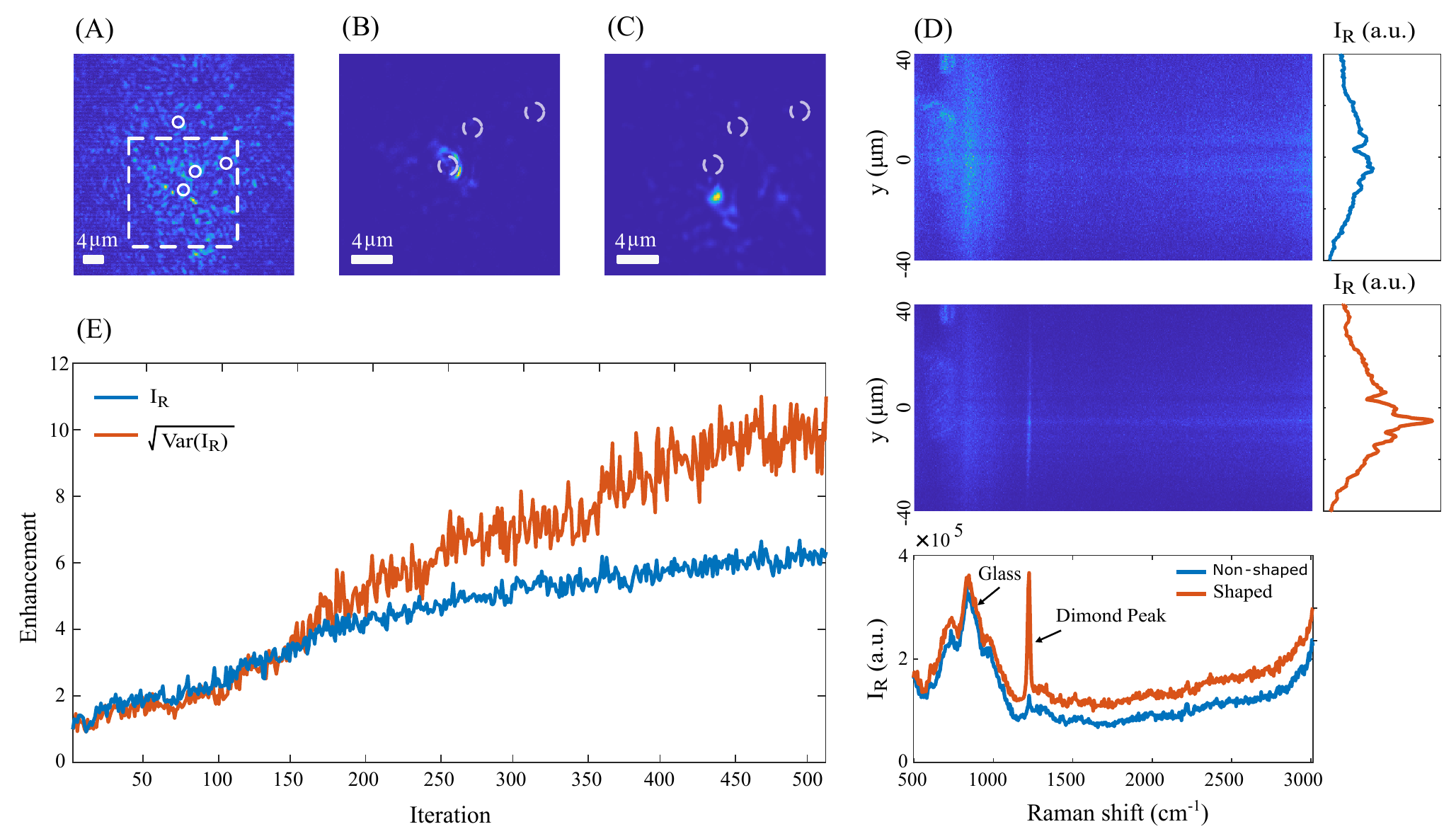}
\caption{
Experimental results. (A) Excitation speckle before wavefront shaping with the four target particles marked by white circles. (B) Shaped excitation focusing on a single particle. The area shown corresponds to the dashed white box in (A). (C) Shaped excitation after shifting the focus off the diamond particle for better visibility. (D) Raman signal recorded by the imaging spectrometer before shaping (top panel) and after shaping (middle panel). The bottom panel shows spatial averages of the unshaped and shaped signal while the right most plots display cuts representing the spatial speckle at the primary Raman shift of diamond. (E) Variance and intensity evolution during the optimization process. Here, we optimize 256 input modes two times each.
}
\label{fig:results}
\end{figure}

Figure~\ref{fig:results} shows results obtained for a sample of four diamond particles where initially  the excitation speckle covers all sources (see Fig.~\ref{fig:results}(A)).
Optimising the incident wavefront to maximize the variance of the one-dimensional Raman speckle we are able to focus on a single target (Fig.~\ref{fig:results}(B)).
Since the diamond particles have a high refractive index they strongly distort the excitation light recorded in transmission, thus preventing the observation of the focus on the control camera right after optimization.
To demonstrate that we actually focus on a single particle only, we slightly move the focus off the diamond using galvanometric mirrors (Fig.~\ref{fig:results}(C)).
For small shifts within the range of the memory effect this does not deteriorate the focus intensity substantially. 
Figure~\ref{fig:results}(D) displays the Raman speckle before and after optimization, showing an enhancement of the diamond signal by a factor $\sim6$ (see Fig.~\ref{fig:results}(E)).
Note that, at the same time the background signal increases by only 30$\%$. 

\begin{figure}[t]
\centering\includegraphics[width=1.0\columnwidth]{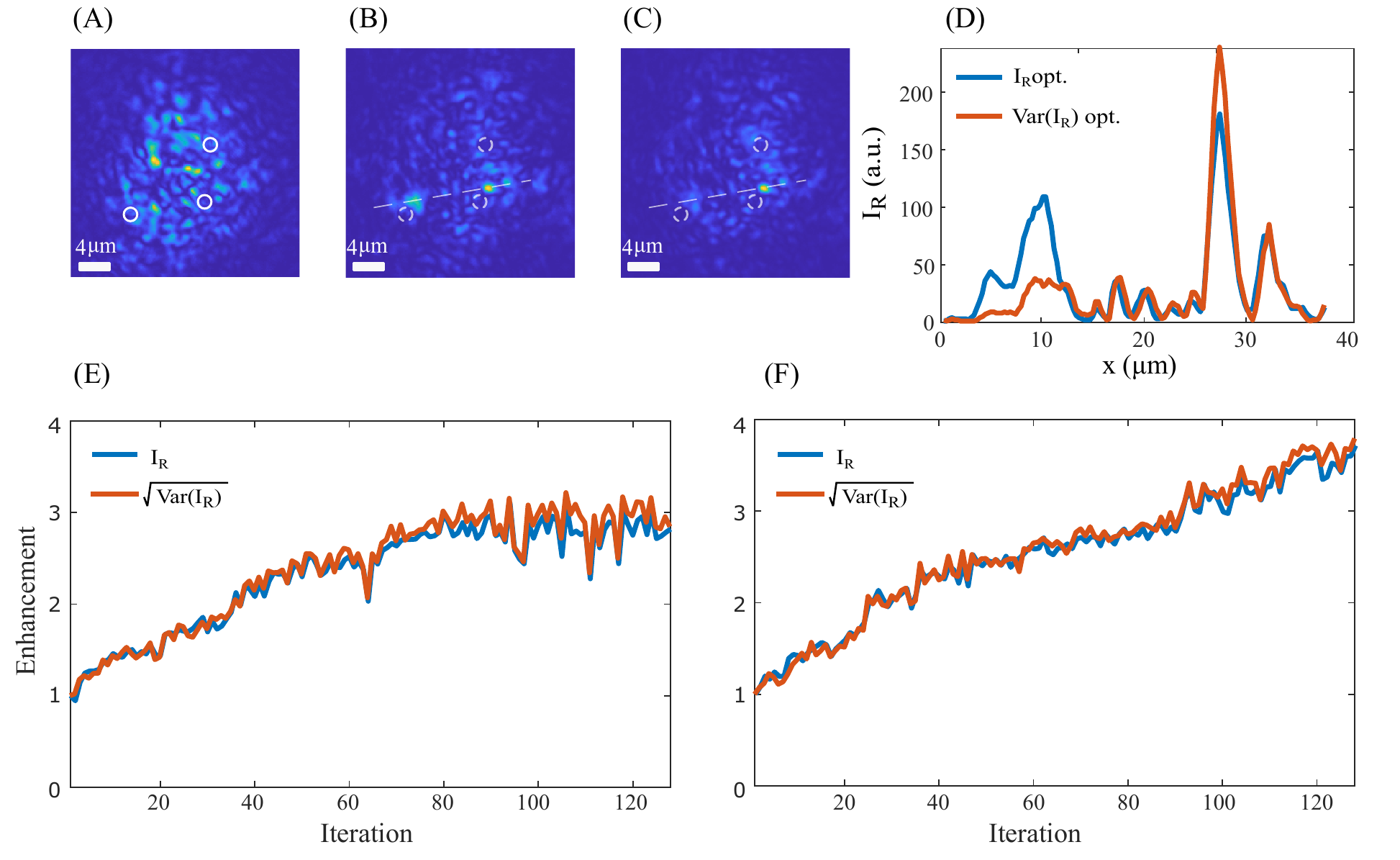}
\caption{Comparison between variance and intensity metrics. (A) Excitation speckle before shaping with the position of three diamond particles marked by white circles. (B) Shaped excitation pattern obtained from an intensity optimization. (C) Same pattern obtained from a variance optimization. For both (B) and (C) the final pattern is slightly shifted in order to analyze the amount of light directed at the targets. (D) Cuts through patterns presented in (B) and (C) as indicated by the dashed white lines. (E) Evolution of the intensity and the square root of the spatial variance of the Raman signal under intensity optimization. (F) Same for the variance optimization. Here, we optimize 64 input modes two times each.}
    \label{fig:comparison}
\end{figure}

To analyze whether the variance metric indeed aids convergence onto a single target, we compare it to the optimization towards a maximal total Raman intensity.
The results are shown in Fig.\ref{fig:comparison}. 
Four particles are illuminated by the initial excitation speckle and both the intensity and the variance optimization manage to concentrate light on the targets (Fig.\ref{fig:comparison}(A-C)). Comparing the final excitation patterns, the variance optimization results in better concentration of power onto a single diamond (Fig.\ref{fig:comparison}(D)).
However, comparing the variance and intensity signals for both optimizations does not show a significant difference (Fig.\ref{fig:comparison}(E-F)).
Here, an unambiguous advantage of the variance metric cannot be shown.

While it is true that in an ideal configuration, like the simulations presented in Fig. \ref{fig:simulation}, intensity optimization does not lead to convergence to a single target, in the experiment its performance is effectively much better.
This is due to several factors.
For one, the finite envelope of the excitation speckle results in an inhomogeneous illumination of the targets, which means the most central target will generate the most signal and optimising towards maximal total intensity will more easily converge there.
In additions, the diamond targets are not equally bright as their size and shape varies.
Another factor is the selection of a 1D slice of the Raman speckle by the imaging spectrometer.
Depending on how the individual targets are positioned relative to this slice their contribution to the signal will change (see SM for details).
All these factors aid convergence for intensity optimizations as they lift the restriction that any distribution of the excitation light on the targets will result in the same signal.

In conclusion, we demonstrate a non-invasive wavefront shaping method that enhances the spontaneous Raman signal after a scattering sample by maximizing its spatial speckle variance.
We achieve focusing on single targets and the largest increase (6 fold) in Raman signal realized so far. 
Furthermore, the method is chemically selective, representing a crucial step towards label-free microscopy within complex samples.
While experimentally we could not unambiguously show the enhanced single particle focusing capabilities of the variance optimization over the intensity optimization, in simulations the advantage is prominent.
Nevertheless, we believe our approach would find immediate impact in systems that generate strong Raman signals (e.g. SERS) with well defined vibrational spectra, as typically found in photothermal therapy SERS probes~\cite{Tabish2020}.

\vspace{5mm}
\noindent{\bf Acknowledgments:} \\
This project was funding by the European Research Council under the grant agreement No. 724473 (SMARTIES), the European Union's Horizon 2020 research and innovation program under the FET-Open grant No. 863203 (Dynamic) and the French National Research Agency (ANR): LabEX ENS-ICFP (ANR-10-LABX-0010/ANR-10-IDEX-0001-02 PSL*). Bingxin Tian is supported by Xi'an Technological University and Bernhard Rauer by the Marie Sk\l odowska-Curie fellowship No.\ 888707 (DEEP3P).

%\newpage
%\nocite{*}
%\bibliographystyle{}
\bibliography{sample}

\clearpage
\onecolumngrid

\renewcommand{\thefigure}{S\arabic{figure}}
\setcounter{figure}{0}

\begin{center}
  \LARGE
  \textbf{Supplementary Materials} %for \\ \textbf{title}
\end{center}

\section{Experimental Setup}
\begin{figure}[htbp]
\centering
\fbox{\includegraphics[width=0.7\linewidth]{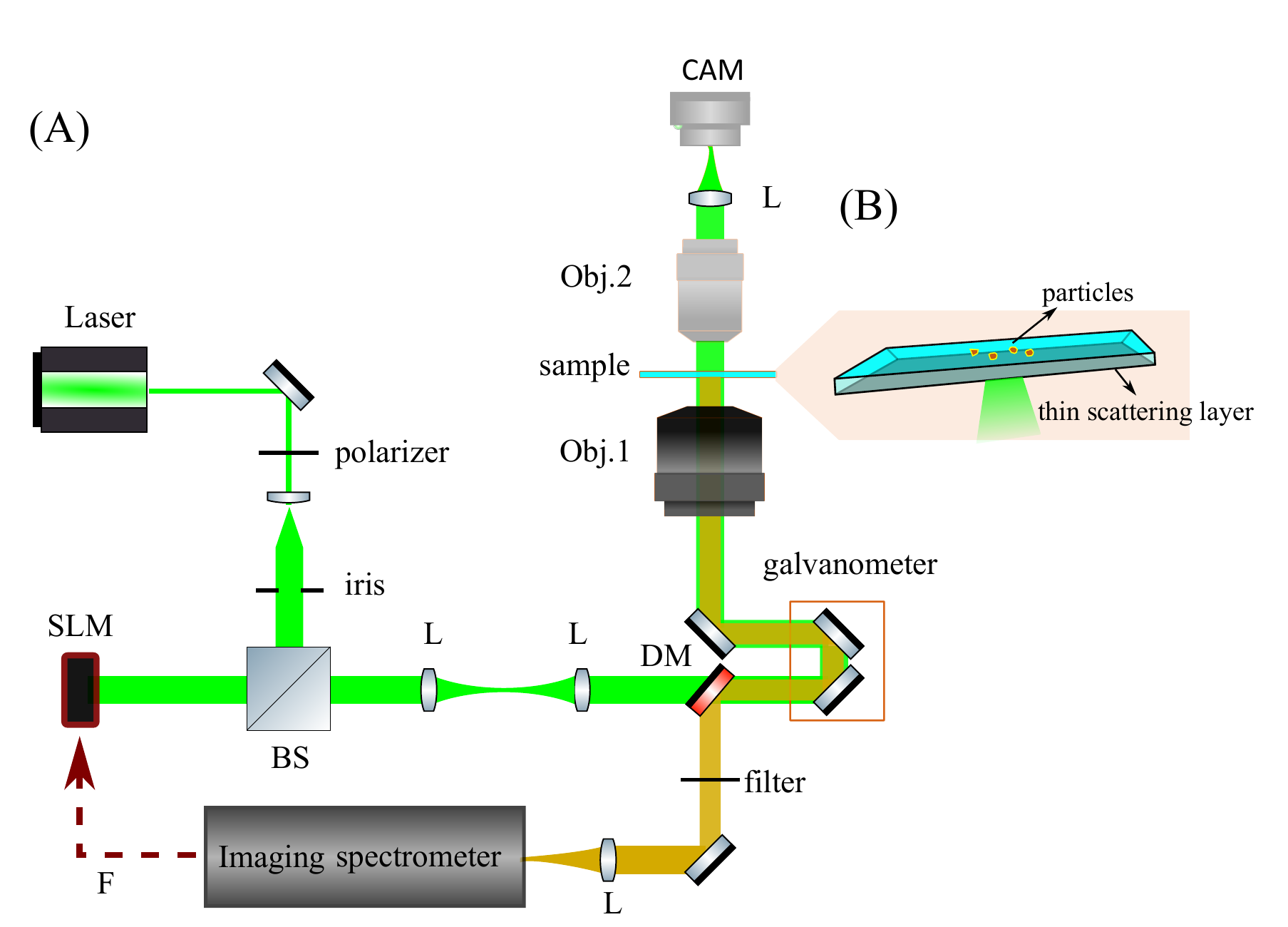}}
\caption{Experimental setup. (A) Schematic of the optical layout. The excitation light is shown in green, the Raman scattered light in orange. SLM: spatial light modulator, L: lens, BS: beam splitter, DM: dichroic mirror, Obj.: objective, CAM: camera. (B) Blow-up of the sample.
}
\label{fig:setup}
\end{figure}

A detailed schematics of the experimental setup is presented in Fig.\ref{fig:setup}.
We use a \SI{532}{\nano\meter} continuous wave laser (Spectra Physics, Millennia) whose wavefront is modulated by a liquid crystal phase-only SLM (Hamamatsu, {LCOS-SLM, 800$\times$600}).
The SLM is conjugated to the back focal plane of the illumination objective (Obj. 1) which focuses the projected pattern onto the sample. 
For the measurements presented in Fig. 3 of the main text, a $40\times$ water immersion objective (Leica, Electrophysiology, MP, Confocal, TL) with an NA of 0.8 was used while for the data presented in Fig. 4, an objective (Nikon, Plan Apochromat VC $60\times /1.4$ Oil)  with an immersion medium consisting of same parts glycerol and water was used.
%\ref{fig:results} \ref{fig:comparison}
The Raman scattered light is collected by the same objective and imaged onto the entrance slit of an imaging spectrometer (Andor Shamrock 303i, equipped with an Andor iXon 897), with a dichroic mirror and an additional notch filter used to block the elastically scattered excitation light.
Typically, an exposure time of \SI{300}{\milli\second} is chosen and six phase steps are measured per mode.
A pair of galvanometeric mirrors is used to shift the excitation pattern illuminating the sample.
In transmission, a second objective (Obj. 2, Olympus
MPlan N 50$\times$/0.75) is used to image the back side of the sample and the excitation light pattern that reaches the targets onto a camera (FLIR, Progressive Scan CMOS).
This transmission imaging system is only used to confirm convergence, it does not play any role in the wavefront optimization process.

\section{Sample preparation}
As a sample we use diamond particles, placed on the top side of a \SI{170}{\micro\meter} thick \#1.5 cover slip (Fig.\ref{fig:setup}(B)).
The particles are extracted from a diamond grinding paste and are roughly \SI{1}{\micro\meter} in size.
The bottom side of the coverslip, facing the illumination objective, was sandblasted to create a rough surface introducing scattering. 
The resulting speckle illuminating the diamond particles shows no ballistic component and we control its envelope by choosing different immersion media or partially covering the sandblasted surface with UV curing optical glue.

\section{Influence of the slit orientation}

\begin{figure}[ht]
\centering
\fbox{\includegraphics[width=0.7\linewidth]{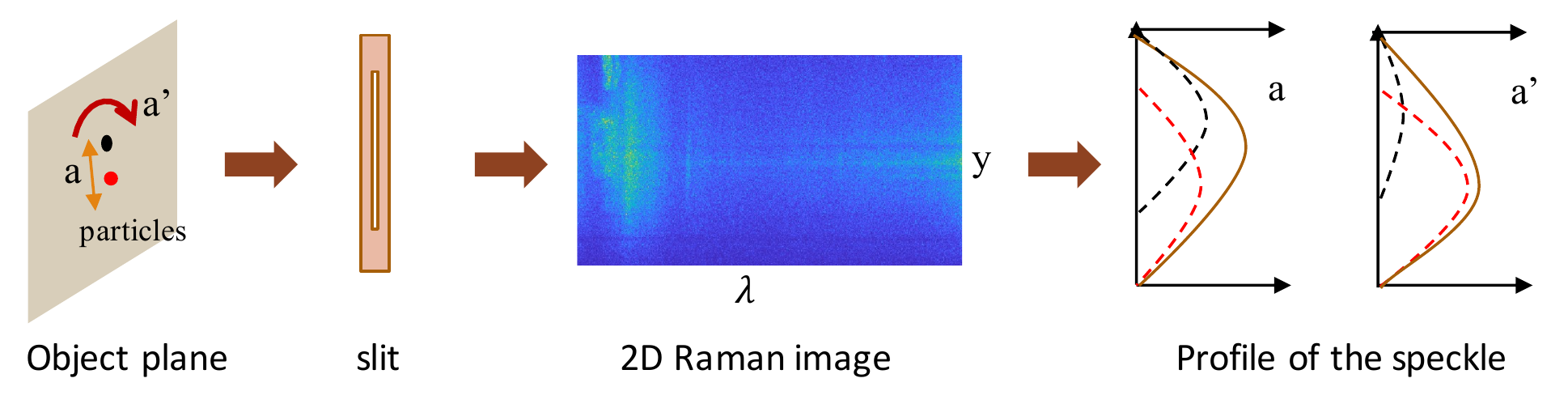}}
\caption{Schematics illustrating the influence of the imaging spectrometer's entrance slit orientation and position with respect to sample.
Let us consider two equally bright particles behind a scattering layer.
Arranged in configuration 'a' they are aligned with the slit orientation and contribute equally to the measured signal.
Arranged in configuration 'a'', however, if the slit position is not aligned exactly with the middle between the two, one particle will contribute stronger to the measured Raman speckle.
}
\label{fig:slit}
\end{figure}

Although the entrance slit of the imaging spectrometer is necessary to measure a contrasted speckle of the Raman scattered light, it also limits the intensity of different target responses, related to their relative position to the slit.
This is because the Raman speckles from particles in different positions are slightly shifted. For the superposed speckle measured by the imaging spectrometer, the position of the slit determines the contribution of each particle, as shown in Fig.\ref{fig:slit}.

\end{document}